\documentclass[12pt]{article}
\usepackage[english]{babel}
\usepackage{amsfonts}
\usepackage{amsmath}
\usepackage{graphicx}
\usepackage{amssymb}

\begin{document}

\begin{center}
{\Large\bf On the non-local hydrodynamic-type system\\
 and its soliton-like solution}
\end{center}

\begin{center}

{\bf\it {V. A. Vladimirov, E. V. Kutafina and B. Zorychta}}
\vspace{5mm}

\vspace{5mm}

{\it Faculty of Applied Mathematics, AGH University of Science and
Technology}

\vspace{5mm}

{\it Mailing address: Mickiewicz Avenue 30, 30-059 Krakow, Poland}

\vspace{5mm}

Email: {\it \underline {vsevolod.vladimir@gmail.com}}

\vspace{5mm}


\end{center}

\vspace{5mm}

{\bf Abstract} { \footnotesize We analyze the  conditions, which
guarantee the existence of  periodic and soliton-like traveling wave
solutions in the non-local hydrodynamic model of structured media.}

\section{Introduction}

In natural science there exists a number of  examples of the
creation and stable evolution  of so called coherent states, or
spatio-temporal patterns \cite{Haken,Schuster,RabTrubeckov}. The
location and specification of such pattens within the confines of
adequate mathematical models is quite difficult, because they are
usually described by non-linear partial differential equations. As
exceptions, the bi-Hamiltonian equations should be mentioned (see
e.g. \cite{olver}, ch. VII), which can be integrated completely by
applying the inverse-scattering method or other associated
procedures \cite{dodd}. Unfortunately, the number of equations which
fall under this method is scarce. And, what is possibly more
important, coherent structures occur very often in open dissipative
systems which cannot be Hamiltonian in principle. In such cases,
only a few approaches, the symmetry-based methods \cite{ovs,olver}
among them, could serve as alternatives to numerical modeling. It is
worth noticing that in recent years additionally, and seemingly
independently to the group methods, the so called anzatz-based
approach is widely used \cite{fan,nikbar,baryur,vladku}. It allows
for solutions with pre-set properties (e.g. soliton-like, kink-like
or periodic stationary  structures or traveling waves) to be
obtained, but, according to our belief, they are less universal than
approaches based on self-similarity methods, supported by
qualitative analysis. This is because in the latter case one can
analyze not only particular self-similar solutions, which
fortunately can be expressed analytically, but the whole family
with the given symmetry.

Let's now focus our attention on a very important factor which is
intensively used in the article. Dissipative models of real
processes very often turn out to be close (to some extent) to
Hamiltonian models. Such a situation occurs for example in cases
when dissipation, relaxation, heterogeneity and similar effects
which ruin the regularity of the model,  display a small-parameter
description. In such a case, while searching for dissipative
structures, not only qualitative analysis can be applied, but also
one can use the whole power of Hamilton's method. Apart from this,
to research a system with small parameters, the ideas and methods of
the concept of approximate symmetry can be used, and physically
justified small modules, regardless of whether or not they break the
symmetry of an unperturbed model, may be taken into consideration.

This article is concerned with the search for wave structures
(particularly soliton-like and periodic regimes) in a hydrodynamic
model, taking into account the effects of spatio-temporal
non-localities. Using the afore-mentioned methods, the criteria for
the existence of the wave patterns will be formulated in a wide set
of parameter values in those cases where the model under
consideration is in some way equivalent to a Hamiltonian one, and in
the case, when dissipative effects (associated with relaxation) and
the mass forces are taken into account.

The structure of the article is as follows. In  section 2 a
hydrodynamic-type model of a non-local medium is formulated, and a
group-theory reduction to the system of ODEs, describing a family of
approximately invariant traveling wave  solutions is performed. In
section 3 the existence of one-parametric family of periodic and
soliton-like solutions is shown when the temporal relaxation and
mass forces are ignored. In section 4, based on a generalization of
Melnikov's method, and  the essential use of the previously obtained
exact solution, the existence of soliton-like regimes is
demonstrated when small  mass force and relaxation effects are
incorporated.

\section{Nonlocal hydrodynamic-type system and its symmetry reduction}

 Below we introduce a modelling system, taking  into account the
non-local effects. These effects are manifested e.g. when an
impulse loading is applied to media, possessing an internal
structure on mesoscale \cite{vsandan1,makar1,peerlart}.
Description of the non-linear waves propagation in such media
depends in essential way on the ratio of  a characteristic size
$\,d\,$ of elements of the medium structure  to a characteristic
length $\lambda$ of the wave pack. If $d\,/\lambda\,$   is $O(1)$,
then the basic concepts of continual media are completely false
and one should use the description, based, e.g. on  the element
dynamics methods \cite{danvengr}.  The classical continuum
mechanics equations applications is justified rather in those
cases, when $d\,/\lambda\,\ll 1,$ and the discreteness of the
matter could be completely ignored.

The models studied in this work apply when the ratio  $d/\lambda$
is much less that unity and therefore the continual approach is
still valid, but it is not  as small that we can ignore the
presence of  the internal structure.
%2
As it have been shown in a number of papers (see e.g.
\cite{vakhkul}), in the long wave approximation  the balance
equations for  mass and momentum retain their classical form, which
in the one-dimensional case can be written as follows:
\begin{equation}\label{Eq:balanceq}
\left\{
\begin{array}{ll}
 u_t+p_x=\Im/\rho , \\
 \rho_t+\rho ^2 u_x =0.
\end{array} \right.
\end{equation}
Here $u$ denotes  mass velocity, $p$ is  pressure, $\rho$ is
density, t is time, $x$ is  mass (lagrangean) coordinate, $\Im$ is
mass force, lower indices denote partial derivatives with respect
to subsequent variables. Thus, the whole information about the
presence of structure in this approximation is contained in a
dynamic equation of state (DES), which should be incorporated to
system (\ref{Eq:balanceq}) in order to make it closed.
\par Generally speaking, DES for multi-component structured media,
manifesting the non-local features,  takes  the form of integral
equation \cite{rudyak,zubti}, linking generalized thermodynamical
flow $J$ and generalized thermodynamical force $X$, causing this
flow:
\begin{equation}\label{Eq:intgen}
J=\int_{-\infty}^{t}\left[\int_{R}K(t,t';\,x,x')X(t',x')\,dx'\right]\,dt'.
\end{equation}
Here $K(t,t';\,x,x')$ is a kernel, taking into account nonlocal
effects. Function $K$ can be calculated, in principle, by solving
dynamic problem  of structure's elements interaction, however such
calculations are extremely difficult. Therefore  in practice one
uses, as a role, some  model kernel, describing well enough the main
properties of  the non-local effects and, in particular, the fact
that these effects vanish rapidly as $|t-t'|$ and $|x-x'|$ grow.
This property could be used in order to pass from the integral
equation (\ref{Eq:intgen}) to a pure differential equation.

% ***************DES*********Spatial non-locality********

One of the simplest state equations  accounting for the effects of
spatial nonlocality takes the form:
\begin{equation}\label{Eq:intspat}
p= \hat{\sigma}\,\int_{-\infty}^{t}K_1(t,\,t')\left[
\int_{-\infty}^{+\infty}K_2\left(x,\,x'
\right)\,\rho^{n}(t',x')\,dx'\right]dt',
\end{equation}
where $ K_2\left(x,\,x' \right)=\exp\left[-{(x-x')^2}/{l^2}\right]
$ (cf. with \cite{peerlart}). Using the fact that the function
$\exp{\left[-(x-x')^2/{l^2}\right]}$ extremely quickly approaches
 zero as $|x-x'|$ grows, we substitute the function  $\rho^{n}(t',x')$
 by the first three terms of its  decomposition into the power series:
\[
\rho^{n}(t',x')=\rho^{n}(t',x)+[\rho^{n}(t',x)]_{x}\frac{x'-x}{1!}+
[\rho^{n}(t',x)]_{xx}\frac{\left(x'-x\right)^2}{2!}+O(|x-x'|^{3}),
\]
obtaining this way the following approximate flow-force relation:
\begin{equation}\label{Eq:tspat}
p= \hat{\sigma}\,\int_{-\infty}^{t}\,K_1\left(t,\,t'\right)\,
L\left(\rho ,\,\rho _x,\,\rho _{xx} \right)dt',
\end{equation}
where
\[
L[\rho,\, \rho_{x},\,\rho_{x\,x}]=
c_{0}\,\rho^{n}(t',x)+c_{2}\,\left[\rho^{n}(t' ,x)\right]_{xx},
\]
\[
c_{0}=l\int_{-\infty}^{+\infty}e^{-\tau^{2}}d\tau=l\sqrt{\pi},
\qquad c_{2}=\frac{l^{3}}{2}\int_{-\infty}^{+\infty}\tau^{2}e^{-\tau^{2}}d\tau=\frac{l^3\,\sqrt{\pi}}{4}.
\]

% ****************Temporal**non-locality******************
Now we are going to analyze different function $K_1\left(t,\,t'
\right)$, responsible for the relaxing effects inside the elements
of the internal structure. A medium with one relaxing component is
usually described by the kernel
\begin{equation}\label{Eq:onerelker}
K_1\left(t,\,t'
\right)=\hat\tau^{-1}\,\exp{\left[-\frac{t-t'}{\hat\tau}\right]}.
\end{equation}
With such a kernel the flow-force relation (\ref{Eq:tspat}) can be
expressed as a first order PDE. In fact, inserting the kernel
(\ref{Eq:onerelker}) into the equation (\ref{Eq:tspat}) and then
differentiating it with respect to $t$, we get the following DES:
\begin{equation}\label{Eq:onerelax}
{\hat\tau}\,p_t+p=\hat\sigma \,L[\rho,\, \rho_{x},\,\rho_{x\,x}].
\end{equation}

Assuming that
\[
K_1\left(t,\,t'
\right)=a_1\,\exp{\left[-\frac{t-t'}{\tau_1}\right]}+a_2\,\exp{\left[-\frac{t-t'}{\tau_2}\right]}
\]
we can obtain the DES for structured media with two relaxing
components. In order to obtain a pure differential flow-force
relation, we should use $p$, $p_t$ and $p_{tt}.$ Taking their linear
combination with properly chosen coefficients, we obtain the
following DES:
\begin{equation}\label{Eq:tworelax}
h\,p_{tt}+\tau\,p_t+p=\alpha\,L[\rho,\,
\rho_{x},\,\rho_{x\,x}]+\,\mu\,L_t[\rho,\,
\rho_{x},\,\rho_{x\,x}],
\end{equation}
where \[
h=\tau_1\,\tau_2,\;\tau=\tau_1+\tau_2,\;\alpha=\hat\sigma\left(a_1\,\tau_1+a_2\,\tau_2\right),\;
\mu=\hat\sigma\,\left(a_1+a_2\right)\tau_1\,\tau_2.
\]

%******************End******Temp*************************

%***********Begin***Oscillation-type**kernel****
It is worth noting that a DES formally identical with
(\ref{Eq:tworelax}) can be obtained from  the equation
 (\ref{Eq:intspat}), using in the temporal part of the  kernel
function
\[
K_1\left(t,\,t'
\right)=\exp{\left[-\frac{t-t'}{\tau_1}\right]}\,\cos{\left[\frac{t-t'}{\tau_2}+\varphi_0\right]},
\]
which contains, beside  the exponentially decaying, the oscillating
term. In fact, using the similar combination as in the previous
case, we get the equation, formally identical with the Eq.
(\ref{Eq:tworelax}):
\begin{equation}\label{Eq:oscrelax}
\bar h\,\hat p_{tt}+\bar \tau\,\hat p_t+\hat p=\,\bar
\alpha\,L[\hat\rho,\,
\hat\rho_{x},\,\hat\rho_{x\,x}]+\,\bar\mu\,L_t[\hat\rho,\,
\hat\rho_{x},\,\hat\rho_{x\,x}],
\end{equation}
with
\[
\bar h=\frac{\tau_1}{2}\,\gamma,\;\bar \tau= \gamma, \;\bar
\alpha=\bar h\,\cos{\varphi_0}\,\gamma, \;\bar
\mu=\frac{\gamma}{2\,\tau_2}\,\left(\tau_1\,\cos{\varphi_0}-\tau_2\,\sin{\varphi_0}\right),
\]
where $\gamma=2\tau_1\,\tau_2^2/\left(\tau_1^2+\tau_2^2\right).$ In
order to maintain the physical meaning of pressure and density we
introduce the variables $\hat p=p-p_0,\;\;\hat\rho=\rho-\rho_0,$
where $p_0>0$ and $\rho_0>0$ are some constant equilibrium values of
the parameters.

 The difference between (\ref{Eq:tworelax}) and
(\ref{Eq:oscrelax}) arises from the fact that the coefficients of
these formally coinciding equations belong to distinct domains of
the parameter space. For example, the coefficients $h$ and $\tau$
from (\ref{Eq:tworelax}) satisfy the relation
\[
4\,h\,<\,\tau^2
\]
provided that $\tau_1 \neq \tau_2$ while for the $\bar   h$ and
$\bar \tau$  the opposite inequality holds. As it is shown in
\cite{vladsid,vladsidskur}, the properties of  traveling wave
solutions to a non-local hydrodynamic-type model depend in essential
way on the values of the parameters.

It is worth noting that DES very similar to (\ref{Eq:tworelax}) have
been obtained  in \cite{vsandan2,dandan} on the basis of the
phenomenological non-equilibrium thermodynamics formalism. We prefer
to present the approach based on the formula (\ref{Eq:intgen}) and
the modeling kernels of non-localities, since it is less cumbersome
than that using the thermodynamical methods.
%********************End OSC**********************************

In this work we concentrate on the study of the system of balance
equations (\ref{Eq:balanceq}), closed by the  DES
(\ref{Eq:onerelax}):
\begin{equation}\label{Eq:mainpde}
\left\{
\begin{array}{ll}
{ u_t}+{p_x}=\Im/\rho , \\
 \rho_{ t}+\rho^{2} u_{ x}=0, \\
\hat\tau p_{
t}+p=\frac{\beta}{\nu+2}\rho^{\nu+2}+\sigma[\rho^{\nu+1} \rho_{x\,x}
+ (\nu+1)\rho^{\nu} \left(\rho_{x}\right)^{2}],\end{array} \right.
\end{equation}
where $\nu =n-2, \quad \beta =c_0\,\hat\sigma \,(\nu +2)\,\hat\tau,
\quad \sigma =c_2\,\hat \sigma \,(\nu +2)\,\hat\tau$.

%************************ VSTAVKA 1 ***********************
In what follows we put in (\ref{Eq:mainpde}) $\hat \tau=\epsilon
\tau,\quad \Im /\rho= \epsilon f(\rho)$ with $|\epsilon|\ll 1.$
Introduction of a small parameter enables us to employ the
approximate symmetry technique and avoid difficulties arising from
the fact that the external force, as well as the relaxing terms, can
destroy the scaling symmetry of the problem.

It is easy to verify  by the straightforward checking, that for
$\epsilon=0$ the system (\ref{Eq:mainpde}) admits the following Lie
symmetry group \cite{olver,ovs} generators:
\begin{equation}\label{Eq:symops}
\begin{array}{ll}
\hat P_0=\frac{\partial}{\partial\,t},\quad \hat
P_1=\frac{\partial}{\partial\,x},  \quad \hat
Z=-\frac{\nu+3}{\nu+1}\,t\,\frac{\partial}{\partial\,t}+u\,\frac{\partial}{\partial\,u}+
\frac{2}{\nu+1}\,\rho\,\frac{\partial}{\partial\,\rho}+\frac{4+2\,\nu}{\nu+1}\,p\,\frac{\partial}{\partial\,p}.
\end{array}
\end{equation}
For small  $\epsilon$, the system (\ref{Eq:mainpde}) admits the
operator
\begin{equation}\label{Eq:approxsym}
\hat X=\hat P_0+D\,P_1+\epsilon\,\xi\hat Z
\end{equation}
in the sense of approximate symmetry \cite{baikov}. Let us note that
the first equation of the system (\ref{Eq:mainpde}) admits the
operator $\hat Z$ in generally accepted sense when
$f(\rho)=a\,\rho^{\nu+2}.$ Nevertheless the set (\ref{Eq:symops}) is
not admitted by the system (\ref{Eq:mainpde}), because of the time
derivative term contained in the DES.

A passage to the self-similar variables enabling to factorize  the
system (\ref{Eq:mainpde}), is based on the operator
(\ref{Eq:approxsym}). The characteristic system corresponding to
this operator, up to $O(\epsilon^2)$, can be presented as follows:
\begin{equation}\label{Eq:charsys}
\left(1+\frac{\nu+3}{\nu+1}\,\epsilon\,\xi\,t\right)\,dt=\frac{d\,x}{D}=
\frac{d\,u}{\epsilon\,\xi\,u}=\frac{d\,\rho}{\frac{2}{\nu+1}\,\epsilon\,\xi\,\rho}=
\frac{d\,p}{\frac{2(2+\nu)}{\nu+1}\,\epsilon\,\xi\,p}.
\end{equation}
Solving the system (\ref{Eq:charsys}), and expressing the initial
variables in terms of its first integrals,  one can construct the
following ansatz:
\begin{equation}\label{Eq:ansatz}
\begin{array}{lll}
u=\left(1+\epsilon\,\xi\,t\right)U(\Omega), &
p=\left(1+\frac{2(2+\nu)}{\nu+1}\,\epsilon\,\xi\,t\right)\Pi(\Omega),&
\rho=\left(1+\frac{2}{\nu+1}\,\epsilon\,\xi\,t\right)R(\Omega),
\end{array}
\end{equation}
where
\[
\Omega=x-D\,t-\frac{\nu+3}{\nu+1}\,\epsilon\,\xi\,t^2\,D/2
\]
 is new independent variable.

The approximate self-similar reduction is performed in several
steps. Inserting ansatz (\ref{Eq:ansatz}) into the second equation
of the system (\ref{Eq:mainpde}), we obtain a first order
differential equation admitting the separation of variables:
\[
R^2\,\dot U-D\,\dot
R+\frac{2}{\nu+1}\,\epsilon\,\xi\,R=O(\epsilon^2).
\]
Integration of this equation gives us the approximate first integral
\begin{equation}\label{Eq:ufirstint}
U(\Omega)=C_1-\frac{D}{R}-\frac{2}{\nu+1}\,\epsilon\,\xi\,\int{\frac{d\,\Omega}{R(\Omega)}}+O(\epsilon^2).
\end{equation}
In what follows we assume that $C_1={D}/{R_1}$, where $0<R_1=
const$. For $\epsilon=0$ such a choice immediately leads to the
asymptotics
\begin{equation}\label{Eq:asymptot1}
\lim_{x\to +\infty}u(t,\,x)=0, \qquad \lim_{x\to
+\infty}\rho(t,\,x)=R_1.
\end{equation}

Inserting ansatz (\ref{Eq:ansatz}) into the first equation of the
system (\ref{Eq:mainpde}),  and using Eq. (\ref{Eq:ufirstint}), we
obtain the equation
\begin{equation}\label{Eq:factorfirst}
\dot \Pi-D\,\left(\frac{D}{R^2}\dot
R-\frac{2}{\nu}+1\,\epsilon\frac{\xi}{R} \right)+\epsilon\,\xi\left(
C_1- \frac{D}{R}\right)-\epsilon\,f(R)=O(\epsilon^2).
\end{equation}
Now we introduce  new  function:
\begin{equation}\label{Eq:G}
G=\Pi+\frac{D^2}{R}.
\end{equation}
Taking derivative of (\ref{Eq:G}) with respect to $\Omega$ and
employing (\ref{Eq:factorfirst}), we obtain the equation
\begin{equation}\label{Eq:eqnforG}
\dot G=\epsilon
\left[f(R)-\xi\,\left(C_1+\frac{1-\nu}{1+\nu}\cdot\frac{D}{R}
\right) \right]+O(\epsilon^2).
\end{equation}

Inserting ansats (\ref{Eq:ansatz}) into the third equation, we get a
second-order ODE. Excluding $\Pi$ from this equation by means of the
formulae (\ref{Eq:factorfirst}), (\ref{Eq:G}), and introducing new
variable $\dot R=Y$ we  finally obtain a closed  system which, up to
$O(\epsilon^2)$, takes the form
\[
\dot R=Y,
\]
\begin{eqnarray}\label{Eq:genODE}
\sigma\,R^{\nu+2}\,\dot Y=
G\,R-\left[D^{2}+\frac{\beta}{\nu+2}R^{\nu+3}+\sigma(\nu+1)R^{\nu+1}Y^{2}\right]-\tau\epsilon\frac{D^3}{R}\,Y,
\end{eqnarray}
\[
\dot G=\epsilon
\left[f(R)-\xi\,\left(C_1+\frac{1-\nu}{1+\nu}\frac{D}{R} \right)
\right]
\]

In the following sections we analyze in detail the system
(\ref{Eq:genODE}) for $\epsilon=0$ and formulate the conditions of
the existence of periodic solutions as well as the homoclinic
regimes, corresponding to soliton-like travelling wave solutions of
the initial system of PDEs.

%************************Vstavka1 END *********************

\section{Qualitative analysis of the reduced system in the case when $\epsilon=0$}

Assuming that $\epsilon=0$, we immediately get that $G=G_0={const}$,
and  the system (\ref{Eq:genODE}) reduces to
\begin{equation}\label{Eq:ds1}
\left\{
\begin{array}{ll}
\frac{dR}{d\omega}=Y \\
\frac{dY}{d\omega}= \left(\sigma R^{\nu+2}\right)^{-1}
\left\{G\,R-\left[D^{2}+\frac{\beta}{\nu+2}R^{\nu+3}+\sigma(\nu+1)R^{\nu+1}Y^{2}\right]\right\},
\end{array} \right.
\end{equation}
where $\omega=x-D\,t$.   Incorporating the conditions
(\ref{Eq:asymptot1}), we express $G_0$ as follows:
\begin{equation}\label{Eq:const_G}
G=\frac{D^{2}}{R_{1}}+\frac{\beta}{\nu+2} R_{1}^{\nu+2},
\end{equation}
%8
Dividing the second equation of the system (\ref{Eq:ds1}) by the
first one, and introducing new variable $Z=Y^2\equiv
(d\,R/d\,\omega)^2,$ we get, after some algebraic manipulation, the
linear inhomogeneous equation
\begin{equation}\label{Eq:lininhom}
{Z'}{(R)}+{2}{[(\nu+1)\,R]^{-1}}Z(R)={2}
\left[G\,R-D^2-{\beta}R^{\nu+3}/{(\nu+2)}
\right]/(\sigma\,R^{\nu+2}).
\end{equation}
Solving this equation with respect to $Z=Z(R)$ and next integrating
the equation obtained after the substitution $Z=(d\,R/d\,\omega)^2$,
we get the following quadrature
\begin{equation}\label{Eq:solds1spat}
\omega-\omega_0=\int\frac{\pm\sqrt{\sigma}\,R^{1+\nu}\,d\,R}{\sqrt{H_1+2\,G\,\frac{R^{2+\nu}}{(2+\nu)}-
2\,D^2\frac{R^{1+\nu}}{(1+\nu)}-\beta\,\frac{R^{2(2+\nu)}}{(2+\nu)^2}}}.
\end{equation}

The direct analysis of the formula (\ref{Eq:solds1spat}) is rather
difficult. To realize what sort of solutions we deal with, the
methods of qualitative analysis can be applied to the  dynamical
system (\ref{Eq:ds1}). It is evident, that all isolated critical
points of the system (\ref{Eq:ds1}) are located on the horizontal
axis $OR$. They are determined by  solutions of the algebraic
equation
\begin{equation}\label{Eq:cp_spat}
P(R)=\frac{\beta}{\nu+2}R^{\nu+3}-GR+D^{2}=0.
\end{equation}
As can be easily seen,  one of the roots of equation
(\ref{Eq:cp_spat}) coincides with $R_{1}$. Location of the second
real root depends on the relations between the parameters. If
 $\nu+3>1$, and $D^2$ satisfies the inequality
\begin{equation}\label{Eq:secroot}
D^{2}>D_{cr}^{2}=\beta R^{\nu+3}_{1},
\end{equation}
then there exists the second critical  point $R_{2}>R_{1}$.
Moreover, if $0 < \nu$ is a natural number, then the polynomial
$P(R)$ has the  representation
\begin{equation}\label{Eq:p_repr}
P(R)=(R-R_{1})(R-R_{2})\Psi (R),
\end{equation}
%9
where
\[
\begin{array}{lc}
\Psi(R)=\frac{\beta\,\left[R^{\nu+1}+R^{\nu}(R_{2}-R_{1})+\dots
+R(R_{2}^{\nu}-R_{1}^{\nu})+(R_{2}^{\nu+1}-R_{1}^{\nu+1})
\right]}{(\nu+2)\left(R_{2}-R_{1}\right)}.
\end{array}
\]
Note that $\Psi(R)$ is positive, for $R>0$. It is a direct
consequence of the  existence of representation (\ref{Eq:p_repr})
valid for any natural $\nu $. But this is also true for any $\nu
>-2,$ or, in other words, whenever the function   $R^{\nu+3}$ is
concave for positive $R$.

Analysis of system's (\ref{Eq:ds1}) linearization matrix
\begin{equation}\label{Eq:mlins}
\hat M(R_i,\,\,0)=\left[
\begin{array}{cc}
0 & 1 \\
\left(\sigma\,R_i^{\nu+2} \right)^{-1}\Psi(R_i)(R_j-R_i) & 0
\end{array}
\right],
\qquad i=1,2, \qquad j\ne i
\end{equation}
shows, that the critical points $A_{1}(R_{1},0)$ is a saddle, while
the critical point  $A_{2}(R_{2},0)$ is a center.  Thus, the system
(\ref{Eq:ds1}) has only such critical points, which are
characteristic to a Hamiltonian system. This circumstance suggests
that there could exist a Hamiltonian system equivalent to the system
(\ref{Eq:ds1}). The Hamiltonian function would help us to make a
complete study of the phase portrait of system (\ref{Eq:ds1}), and
find out  homoclinic trajectories, which correspond to a
soliton-like wave packs, providing that such trajectories do exist.

%11
If we introduce a new independent variable $T,$ obeying the equation
$ \frac{d}{dT}=\sigma R^{\nu+2}\phi(R,Y)\frac{d}{d\omega}, $ then we
can write down the system (\ref{Eq:ds1}) as follows:
\begin{equation}\label{Eq:ds1h}
\left\{
\begin{array}{ll}
 \frac{dR}{dT}=\sigma R^{\nu+2}\,Y\,\phi (R,Y), \\
\frac{dY}{dT}= \phi (R,Y)
\left(GR-\left[D^{2}+\frac{\beta}{\nu+2}R^{\nu+3}+
\sigma(\nu+1)R^{\nu+1}Y^{2}\right]\right). \end{array}\right.
\end{equation}
Here $\phi(R,Y)$ is a function, that is to be chosen in such a way
that the system (\ref{Eq:ds1h}) be Hamiltonian. So we are looking
for a function $H(R,Y)$, satisfying the system
%12
\[
%\begin{equation}\label{Eq:haux1}
%\left\{
%\begin{array}{ll}
\frac{\partial H}{\partial Y}=\sigma R^{\nu+2}\phi (R,Y), \]
\[
\frac{\partial H}{\partial R}= -\phi (R,Y)
\left\{GR-\left[D^{2}+\frac{\beta}{\nu+2}R^{\nu+3}+\sigma(\nu+1)R^{\nu+1}Y^{2}\right]\right\}.
%\end{array}
%\right.
%\end{equation}
\]
Equating mixed derivatives of $H,$ we obtain  the characteristic system
\begin{equation}\label{Eq:multip}
\frac{dR}{\sigma R^{\nu+2}Y}=
\frac{dY}{GR-[D^{2}+\frac{\beta}{\nu+2}R^{\nu+3}+
\sigma(\nu+1)R^{\nu+1}Y^{2}]}=\frac{d\phi}{\nu\sigma Y
R^{\nu+1}\phi}.\end{equation}

The general solution of  system (\ref{Eq:multip}) can be presented
in the form
\[
\phi=R^{\nu}\Phi(\rho),
\]
where $\Phi(\cdot)$ is an arbitrary function of the variable
%13
 \[
\rho=R^{\nu+1}\left\{Y^2\Phi+{D^{2}}{\sigma}^{-1}R^{-(\nu+1)}\log\,R+
R^{2}{\beta}/[{\sigma(\nu+2)(\nu+3)]}-{G}{\sigma}^{-1}R^{-\nu}\right\}.
\]
Putting $\phi=2R^{\nu},$ we can easily restore the Hamiltonian
function:
\begin{equation}\label{Eq:hfun1}
 H=2D^{2}\frac{R^{\nu+1}}{\nu+1}+\frac{\beta}{(\nu+2)^{2}}R^{2(\nu+2)}+
 \sigma Y^{2}R^{2(\nu+1)}-2G\frac{R^{\nu+2}}{\nu+2}.
\end{equation}

As is well-known, the function $H$ is constant on  the phase
trajectories of the both system (\ref{Eq:ds1}) and (\ref{Eq:ds1h}),
and since the integrating multiplier $\phi=2R^{\nu},$ appearing in
the formula (\ref{Eq:ds1h}) is positive for $R>0$, then the phase
portraits of the systems (\ref{Eq:ds1}) and (\ref{Eq:ds1h})
geometrically are identical in the right part of the phase plane
$(R,\,Y)$. Thus, all the statements concerning the geometry of the
phase trajectories of the system (\ref{Eq:ds1h}) lying in the right
half-plane is is valid to the corresponding  solutions of the system
(\ref{Eq:ds1}).

Using the linear analysis, we have already shown that the critical
point $A_{2}(R_{2},\,\,0)$ is  a center. Stated above relations
between systems (\ref{Eq:ds1}) and (\ref{Eq:ds1h}) enable to
conclude that this point does not change when the nonlinear terms
are added. This means that the critical point $A_{2}(R_{2},\,\,0)$
is surrounded by the  closed trajectories, and, hence, the
unperturbed source system (\ref{Eq:mainpde}) possesses a
one-parameter family of periodic solutions. If the right branches of
the separatrices of the saddle point  $A_{1}(R_{1},\,\,0)$ go to
infinity (the stable branch $W^{s}$ when $t\to-\infty$ and the
unstable branch $W^{u}$ when $t\to+\infty$), then the domain of
finite periodic motions is unlimited. Another possibility is
connected with the existence of the  limiting trajectory,
bi-asymptotic to the saddle.  In this case the domain of periodic
solutions is bounded, and the source system, in addition to a
one-parametric family of periodic solutions, possesses localized
soliton-like regimes, corresponding to the homoclinic loop.
%%16
To answer the question on which of the above mentioned possibilities
takes place, the behavior of the saddle separatices, lying to the
right from the line $R=R_1$ should be analyzed.
 We obtain the equation for saddle separatices   by putting
$H=H(R_{1},0)=H_{1}$ in the left hand side of the equation
(\ref{Eq:hfun1}), and solving it with respect to  $Y$:
 \begin{equation}\label{Eq:separ1}
Y=\pm \frac{\sqrt{H_{1}+2G\frac{R^{\nu+2}}{\nu+2}-
[2D^{2}\frac{R^{\nu+1}}{\nu+1}+\frac{\beta}{(\nu+2)^{2}}R^{2(\nu+2)}]}}{\sqrt{\sigma
}R^{\nu+1}}\end{equation} It is evident from equation
(\ref{Eq:separ1}), that incoming and outgoing separatrices are
symmetrical with respect to  $OR$ axis. Therefore  we can restrict
our analysis, e.g., to the upper separatrix $Y_{+}$. First of all,
let us note, that  separatrix $Y_{+}$ forms a positive angle with
the $OR$ axis in the point$\left(R_1,\,\,0 \right)$:
\[
\alpha= \arctan\,\sqrt{{(R_{2}-R_{1})\Psi (R_{1})}/\left({\sigma
R_{1}^{\nu+2}}\right)}.
\]
The above formula arises from the linear analysis of the system
(\ref{Eq:ds1}) in critical point $A_{1}(R_{1},\,\,0)$. So $Y_{+}(R)$
is increasing  when $R-R_{1}$ is small and positive. On the other
hand,  the function
\[
Q(R)={H_{1}+2G\frac{R^{\nu+2}}{\nu+2}-
\left[2D^{2}\frac{R^{\nu+1}}{\nu+1}+\frac{\beta}{(\nu+2)^{2}}R^{2(\nu+2)}\right]},
\]
appearing in the RHS of the formula (\ref{Eq:separ1}), tends to
$-\infty$ as $ R\to +\infty$, because the coefficient at the highest
order monomial $R^{2(\nu+2)}$ is negative, while the index $\nu+2$
is assumed to be positive. Therefore the function $Q(R)$ intersects
the open set $R>R_1$ of the $OR$ axis at least once. Let us denote
the point of the first intersection  by $R_{3}$, and  let us assume
that $R_{3} > R_{2}.$ If, with this assumption, we were able to
prove that $Y_{\pm}(R)$ form the right angle with the $OR$ axis at
the point $R_{3}$, then we would have the evidence of tangent
intersection of the stable and unstable saddle separatrices.

We begin with the note that $\lim_{R\to R_{3}^{-}}Q(R)=+0$.
Calculating derivative  of $Q(R)$ we get:
\begin{equation}\label{Eq:gprime}
Q^{\prime}(R)=-2R^{\nu}\left(\frac{\beta}{\nu+2}R^{\nu+3}-GR+D^{2}\right)=-2R^{\nu}P(R).
\end{equation}
It follows from the decomposition (\ref{Eq:p_repr}), that
$Q^{\prime}(R)<0$ when $R>R_{2}$.   Therefore
\[
\lim_{R\to
R_{3}^{-}}\frac{dY}{dR}=\frac{RQ'(R)-2(\nu+1)Q(R)}{2\sqrt{\sigma
Q}R^{\nu+2}}=-\infty.
\]
So, to complete the prof, we must show that the inequality
$R_{3}>R_{2}$ is true.
%19
 Supposing that the inequalities $R_{1}<R_{3}<R_{2}$ take place, we obtain from the equations
(\ref{Eq:gprime}), (\ref{Eq:p_repr}) that   $ \lim_{R\to
R_3^{-}}{Y_{+}^{\prime}(R)}= +\infty$. On the other hand, the
function $Y_{+}(R)$  approaches zero remaining positive as $R\to
R_{3}^{-}$. But such behavior  is impossible for any function, being
regular inside the interval $\left(R_1,\,\,R_{3}\right)$. The case
$R_{3}=R_{2}$  should also be excluded, because the critical point
$A_{2}(R_{2},0)$ is a center. The result obtained can be formulated
as follows.

{\bf Theorem.} {\it If $\nu>-2$ and $D^2>\beta\,R_1^{\nu+3}$, then
the system (\ref{Eq:ds1}) possesses a one parameter family of
periodic solutions, localized around the critical point
$A_2\left(R_2,\,\,0 \right)$ in a bounded set $\mathbf M.$ The
boundary of this set is formed by the homoclinic intersection of
separatrices of the saddle point $A_1\left(R_1,\,\,0 \right)$. }

\par Thus the unperturbed source system (\ref{Eq:mainpde}) possesses periodic and soliton-like
invariant solutions. Let us note in conclusion that for some special
case the integral at the RHS of the formula (\ref{Eq:solds1spat})
or, what is the same, at the RHS of the formula
\begin{equation}\label{Eq:solds1HMLT}
T-T_0=\pm
\int\frac{d\,R}{2\,\sqrt{\sigma}\,R^{1+\nu}\,\sqrt{H_1+2\,G\,\frac{R^{2+\nu}}{(2+\nu)}-
2\,D^2\frac{R^{1+\nu}}{(1+\nu)}-\beta\,\frac{R^{2(2+\nu)}}{(2+\nu)^2}}},
\end{equation}
can be calculated explicitly:
\begin{equation}\label{Eq:exactHCL}
T=\pm\frac{1}{7\,\sqrt{2}}\Biggl\{{7}\,\ln\left[\frac{R-1} {3-R+
\sqrt{7-2\,R-R^2}}\right]+2\,\sqrt{7}\ln\left[\frac{7-R+\sqrt{7}\sqrt{7-2\,R-R^2}}{R}\right]\Biggr\}.
\end{equation}
This solution corresponds to the following values of the parameters:
$D=1=R_1=\sigma=1,$ $\beta=1/2,$ $\nu=0$, $G=5/4$, and $T_0=0.$
%**********PERIODIC********************************
%**********PERIODIC********************************
%**********PERIODIC********************************

%************************Vstavka2 begin *********************

\section{Homoclinic solutions of perturbed system}

The purpose of this section is to analyze whether and when the
perturbed system (\ref{Eq:genODE}) possesses the homoclinic
solution. Our analysis is based on the generalized Melnikov theory
presented in \cite{wig}. The theory is based in essential way on
the knowledge of the homoclinic solution of unperturbed system,
which enables to measure the distance between the stable and
unstable saddle separatrices for small $\epsilon$. For this reason
we specify the parameters in such a way that they fit the exact
solution (\ref{Eq:exactHCL}).

Since we are going to use the formalism developed in \cite{wig}, we
should pass in (\ref{Eq:genODE}) to the independent variable $T$,
for which the unperturbed system becomes Hamiltonian. The standard
representation for our system in this case will be as follows:
\begin{eqnarray}\label{Eq:pertHDS}
%\left\{
&\frac{d\,X}{d\,T}=J\,D_X\,H( X,\,G)+\epsilon F( X,\,G), \nonumber\\
&\frac{d\,G}{d\,T}=\epsilon L( X,\,G),
%\right.
 \label{eq:4.1.1e}
\end{eqnarray}
\noindent where $ X=colon\,(R,\,Y),$
$D_X=colon\left\{\frac{\partial}{\partial\,R},\,\,\frac{\partial}{\partial\,Y}
\right\}$, $0<\epsilon\ll 1,$ \[ F(X,\,G)=colon(-2\,D^3\,Y/R,\,0),
\]
\[
L(X,\,G)=2\,\left[f(R)-\xi\,\left(C_1+\frac{1-\nu}{1+\nu}\cdot
\frac{D}{R} \right) \right]
\]
while \[J=\begin{bmatrix}
  0 & 1 \\
  -1 & 0 \\
\end{bmatrix}.\]
\noindent In what follows we put $f(R)=a\,R^2+b.$

So we are going to measure the distance between the saddle
separatrices of the perturbed system (\ref{Eq:pertHDS}). The problem
we deal with differs from the classical one \cite{meln,G-H} since in
the unperturbed case the Hamiltonian (\ref{Eq:hfun1}) depends on an
extra variable $G$, which plays the role of a parameter. We assume
that $G$ belongs to an open set $U$, containing
$G_0=D^2/R_1+\beta\,R_1^{\nu+2}/(\nu+2)$.
%So we have the geometric
%picture shown in Fig. 1.
%*********************************
%\begin{figure}
%\end{figure}
%  Fig. 1
%***********************************

Following \cite{wig}, we introduce the set
\begin{multline}
\mathcal{M}=\{ ( X,\,G)\in \mathbb{R}^{2}\times\mathbb{R}^1:\,
X=\gamma(G),\;
\text{where }\,\gamma(G)\,\text{ solves }\\
D_X\,H(\gamma(G),\,G)=0 \text{ and }\det [D_X^2H(\gamma(G),G)]\neq
0,\forall G\in U\}.\label{eq:4.1.11}
\end{multline}
We denote  by $W^s(\mathcal{M})$ and $W^u(\mathcal{M})$ the
$1+1$-dimensional stable and unstable manifolds of $\mathcal{M}$. It
is obvious that the manifolds $W^s(\mathcal{M})$ and
$W^u(\mathcal{M})$ coincide as $\epsilon = 0$ (in this case
$W^s(\mathcal{M})=W^u(\mathcal{M})=\Gamma$).  For sufficiently small
$\epsilon$,  $\mathcal{M}$ is transformed into the locally invariant
manifold $W^{\varepsilon}(\mathcal{M})$. Up to $O(\varepsilon^2)$,
the dynamics on $W^{\varepsilon}(\mathcal{M})$ is governed by the
equation
\begin{equation}\label{Eq:Gonmanif}
\dot{G}=\varepsilon L(\gamma(G),G).
\end{equation}
 For $b=2\,\xi-a\,\,$, $\,W^{\epsilon}(\mathcal{M})$ possesses  the
 equilibrium point $\left(1,\,0,\,5/4 \right).$  Analysis of the
 linear part of equation (\ref{Eq:Gonmanif}) shows, that this point is
 stable as $a>\xi/2$ and unstable as $a<\xi/2$.

 For nonzero $\epsilon$, the invariant manifolds
 $W^s(\mathcal{M})$ and $W^u(\mathcal{M})$ are transformed into
 the (locally) invariant manifolds $W^s_\epsilon(\mathcal{M})$ and
 $W^u_\epsilon(\mathcal{M})$, which do not coincide. Yet at
 certain conditions the manifolds can still have the points of
 intersection, different from $W^{\varepsilon}(\mathcal{M})$. To
 state these conditions, a distance separating
 $W^s_\epsilon(\mathcal{M})$ and $W^u_\epsilon(\mathcal{M})$ is
 measured in vicinity of a point $p\in \Gamma$, lying in the
 section $G_1=5/4$.
 %(see Fig. 2).
 Up to $O(\varepsilon)$, this
 distance is equal to the Melnikov integral (\cite{wig}, Ch. IV):
\begin{multline}\label{Eq:Melnint}
M^{{G_1}}(a,\xi)=\int_{\Gamma(G)}\Bigl[\langle D_{X}H,F{(X,G)}
\rangle +\\+\langle D_{X}H,(D_GJD_{X}H)\int L(X,\,G) dT\rangle
\Bigr](X^{G_1}(T),G_1,T)dT,
\end{multline}
where $X^{G_1}(T)=(R^{G_1}(T),Y^{G_1}(T))$ is the homoclinic
trajectory of the unperturbed system on the $G=G_1$ level
corresponding to the hyperbolic fixed point of the vector field on
$\mathcal{M}_{\varepsilon}.$ The dependence on
$(X^{G_1}(T),G_1,T)$ means in other words that the integration is
carried out along the unperturbed trajectory $\Gamma$. Formula
(\ref{Eq:exactHCL}) describes this trajectory as an implicit
function $T=T(R^{G_1}).$

Let us begin the calculation of Melnikov integral. Below we
present in explicit form the terms from the expression
(\ref{Eq:Melnint}):
\begin{equation*}
\begin{split}
&D_{X}H=\begin{bmatrix}
  2Y^2R+2+\frac{1}{2}R^3-2R\,G \\
  2YR^2 \\
\end{bmatrix},\\
&JD_{X}H=\begin{bmatrix}
  0 & 1 \\
  -1 & 0 \\
\end{bmatrix}, (D_{X}H)=\begin{bmatrix}
  2YR^2 \\
  -2Y^2R-2-\frac{1}{2}R^3+2R\,G  \\
\end{bmatrix},\\
&D_GJD_{X}H=\begin{bmatrix}
  0  \\
  2R \\
\end{bmatrix},
\end{split}
\end{equation*}
\begin{equation*}
\int L \,dT=\int
2R^2[a(R^{G_1}(t))^2+\xi-a-\frac{\xi}{R^{G_1}(t)}]dT.
\end{equation*}
Let's  replace  $T$ with (\ref{Eq:exactHCL}). Then:
\begin{equation*}
dT=\frac{1}{2R^2}\sqrt{\frac{R^2}{\frac{7}{8}-2R-\frac{1}{8}R^4+\frac{5}{4}R^2}}dR.
\end{equation*}
Finally
\begin{eqnarray*}
\begin{split}
&\int L\, dT= \int
(aR^2+\xi-a-\frac{\xi}{R})\sqrt{\frac{R^2}{\frac{7}{8}-2R-
\frac{1}{8}R^4+\frac{5}{4}R^2}}dR=\\
&= \sqrt{2}a\left[1-R^{G_1}(T)\right]\sqrt{7-2R^{G_1}(T)-
(R^{G_1}(T))^2}+
\\
&\qquad\qquad+2\sqrt{2}(\xi+4a)\arcsin\left[\frac{R^{G_1}(T)+1}{2\sqrt{2}}\right].
\end{split}
\end{eqnarray*}
So the Melnikov integral will be of the form:
\[
M^{G_1}(a,\xi)=I_2-I_1,
\]
where
\[
I_1=\underset{\Gamma(G)}\int 4 R^{G_1}(T)
\left[Y^{G_1}(T)\right]^2\,dT,
\]
\begin{eqnarray*}
I_2=8\sqrt{2}\underset{\Gamma(G)}{\int}R^{G_1}(T)
\left[R^{G_1}(T)\right]^3Y^{G_1}(T)\Bigl\{\frac{a}{2}\left[1-R^{G_1}(T)
\right]\sqrt{6-\left[1+R^{G_1}(T) \right]^2}+  \Bigr. \\
\Bigl.
\left(\xi+4\,a\right)\,\arcsin{\frac{R^{G_1}(T)+1}{2\sqrt{2}}}\Bigr\}
\end{eqnarray*}
where $\Gamma(G_1)$ is the unperturbed homoclinic trajectory on the
$G_1$ level.

  The computation of $I_1$ is based on the Green's theorem. As
 \[Y^{G_1}(T)=\frac{1}{2(R^{G_1}(T))^2}\frac{dR^{G_1}(T)}{dT}\;,
 \]
 we  have:
 \[
-I_1=\int_{\Gamma(G_1)}-4(Y^{G_1}(T))^2R^{G_1}(T)dT=\oint_{\Gamma(G_1)}-2\frac{Y}{R}dR+0dY
\]
Next,
\[
\oint_{\Gamma}-2\frac{Y}{R}dR+0dY= \int_{1}^{2\sqrt{2}-1}
\int_{-\frac{(R-1)\sqrt{7-2R-R^2}}{2\sqrt{2}R}}^{\frac{(R-1)
\sqrt{7-2R-R^2}}{2\sqrt{2}R}}\quad\frac{2}{R}\,dYdR=
\]
\[
=\int_{1}^{2\sqrt{2}-1}\frac{2(R-1)\sqrt{7-2R-R^2}}{\sqrt{2}R^2}dR=
\frac{4}{7}\sqrt{2}(-7+\sqrt{7}\ln (8+3\sqrt{7}))=0,262789.
\]
On calculating the $I_2$, we use the fact that
$Y^{G_1}(T)={2(R^{G_1}(T))^{-2}} \frac{dR^{G_1}(T)}{dT}:$
\begin{equation}
\begin{array}{ll}
&\int_{\Gamma(G_1)}8\sqrt{2}(R^{G_1}(T))^3Y^{G_1}(T)(-\frac{1}{2}a(R^{G_1}(T)-1)
\sqrt{7-2R^{G_1}(T)-\left[R^{G_1}(T)\right]^2}
+\\
\\
&+(\xi+4a)\arcsin(\frac{R^{G_1}(T)+1}{2\sqrt{2}}))dT=\\
\\
&=2\,\int_{1}^{2\sqrt{2}-1}4\sqrt{2}R(-\frac{1}{2}a(R-1)\sqrt{7-2R-R^2}
+(\xi+4a)\arcsin(\frac{R+1}{2\sqrt{2}}))dR=\\
\\
&=-8\sqrt{2}(-3+\pi)a+(4\sqrt{2}+(-8+ 6\sqrt{2})\pi)(\xi+4a)=\\
\\
&=2\left[-1,60194a+7,18141(\xi+4a)\right].
\end{array}
\end{equation}
So, finally
\begin{equation}
\label{eq:mel} M^{G_1}(a,\xi)=0,262789-2\cdot\left[
1.60194a+7,18141(\xi+4a)\right],
\end{equation}
amd the Melnikov's integral vanishes if
\begin{equation}\label{eq:locusHCL} \xi=-0,03659-3,7769a.
\end{equation}
Then, for $\epsilon$ sufficiently small, stable and unstable
manifold intersect near the values of  $({a},\,{\xi})$, defined by
equation (\ref{eq:locusHCL}).

\begin{center}

\begin{tabular}{|c||c|c|c|c|c|c|c|c|c|c|c|c|c|} \hline
a  & -0.09 & -0.07 & -0.05 & -0.04 & -0.025 & -0.01  \\
\hline $\delta$   & -0.0214   & -0.0186  & -0.01567  & -0.014175 &
-0.0115475 & -0.009478  \\ \hline \hline
a &  0.01 & 0.025 & 0.04 & 0.06 & 0.08 & 0.1 \\
\hline $\delta$  & -0.0065 & -0.0041 & -0.0017 & 0.000125 & 0.0052 &
0.0089  \\ \hline

\end{tabular}
\end{center}

\begin{figure}
\begin{center}
\includegraphics[width=3.in, height=2.25 in]{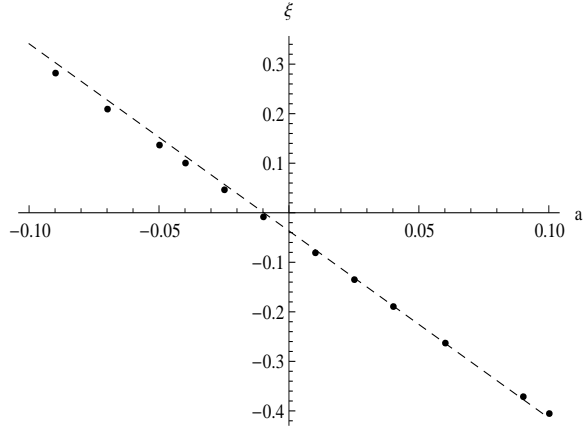}
\caption{The points $({a},\,{\xi})$, corresponding to the homoclinic
intersection (boldface dots) on the background of the straight line
(\ref{eq:locusHCL}) }\label{Fig:lociHCL}
\end{center}
\end{figure}

The results obtained was verified by direct numerical simulation. To
gain the locus of the points in the parameter plane $({a},\,{\xi})$
corresponding to the homoclinic intersection, we used the following
modification of the approximate formula (\ref{eq:locusHCL}) obtained
with the help of the Melnikov method:
\[
\xi=-0,03659-3,7769\,a+\delta.
\]
The values of the parameters at which the homoclinic intersection is
observed are presented in the table shown above. The dynamical
system (\ref{Eq:genODE}) was numerically integrated under the
following values of the parameters: $\epsilon=-0.001$,
$D=1=R_1=\sigma=1,$ $\beta=1/2,$ $\nu=0$, $G=5/4$.  The points of
the parametric plane $({a},\,{\xi})$ corresponding to the homoclinic
intersection are shown in figure~\ref{Fig:lociHCL}, together with
the straight line given by the formula (\ref{eq:locusHCL}).

%*********************
%********
\section{Concluding remarks}

We have considered  a non-local hydrodynamic-type model of
structured media  and studied a family of traveling wave (TW)
solutions. Our analysis shows that in the case of a pure spatial
non-locality and the absence of mass forces a reduced ODE system is
equivalent to a Hamiltonian one. It is worth noting that the initial
system of PDEs is not Hamiltonian for the  physically justified
values of the parameters.

Equivalence of the reduced system to a Hamiltonian one is employed
in this work to obtain rigorous proof of the existence of a
one--parameter family of periodic and soliton-like TW solutions. At
certain values of the parameters these solutions possess an analytic
representation, while in general they are described in terms of
Jacobi elliptic functions.

Having established a soliton-like solution to the rigorously
reducible model, it is possible to analyze the soliton-like
solutions' existence in general case, when the exact symmetry of the
modeling system has broken down. Using, in this case, the ideas of
the approximate symmetry and the generalized Melnikov method, we've
stated in explicit form the conditions of the saddle separatrices
intersection in cases when the small mass force and the effect of
temporal non-locality are incorporated.  As there arise from the
general considerations presented in \cite{wig}, intersections of
stable and unstable saddle separatrices are not transversal for the
given type of perturbation and this is backed by the results of our
numerical experiments, that enable us to reveal the loci of the
saddle separatrices intersection with no signs of a "homoclinic
blow-up" observed.

In view of this, it is interesting to investigate what type of
temporal non-locality and external force would lead to the
transversal intersection and, as a consequence, to  drastic changes
in the phase portrait of the reduced system. Our preliminary
analysis shows that the transversal intersection can take place in
the case of periodic mass force and the DES (\ref{Eq:tworelax}),
describing the structured media with two relaxing processes. In
fact, the second relaxing process incorporation causes the dynamic
equation of state to include as a singular perturbation the second
derivative with respect to time. An extra additional variable,
playing the same role as $G$,  will then appear in the dynamical
system obtained via the approximate symmetry reduction. Let us note
in conclusion, that the role of spatial non-locality on the wave
patterns' formation and evolution have been discussed in
\cite{dandan}. In particular it was shown in this work  that chaotic
wave patterns can be gained  via homoclinic bifurcation in the
hydrodynamic-type model accounting for the effects of spatial
non-localities and where, as in the DES (\ref{Eq:tworelax}), the
term $p_{tt}$ is incorporated as a small addend.

%******

%***

\begin{thebibliography}{99}
\bibitem{Haken}
H. Haken, {\em Synergetics. An Introduction}, Springer, New York,
1979.
\bibitem{Schuster}
H. Schuster, {\em Deterministic Chaos}, Wiely, New York, 2005.
\bibitem{RabTrubeckov}
M. Rabinovich, D. Trubeckov, {\em An Introduction in the Theory of
vibrations and Waves}, Nauka Publ., Moscow, 1984 (in Russian).
\bibitem{olver} P.Olver: {\em Applications of Lie Groups to Differential Equations\/},
Springer, New York, 1993.
\bibitem{dodd}
R.K.Dodd, J.C.Eilbek, J.D.Gibbon, H.C.Morris: {\em Solitons and
Nonlinear Wave Equations\/}, Academic Press, London 1984.
\bibitem{ovs} L.V. Ovsiannikov: {\em Group Analysis of Differential Equations\/},
Academic Press, New York, 1982.
\bibitem{fan}
E.Fan, {\em Journ of Physics A: Math and Gen.\/}, {\bf  35},
6853--6872 (2002).
\bibitem{nikbar}
A. Nikitin, T.  Barannyk, {\em Solitary Waves and Other Solutions
for Nonlinear Heat Equations\/}, arXiv:math-ph/0303004 (2003).
\bibitem{baryur}
A.Barannyk, I.Yurik, {\em Proceedings of Institute of Mathematics of
NAS of Ukraine}, {\bf 50(I)}, 29--33 (2004).
\bibitem{vladku}
V.A. Vladimirov and E.V. Kutafina,   {\em Rep.  Math. Physics\/},
{\bf  54}, 261--271  (2004).
\bibitem{vsandan1}
V. Danylenko, V. Sorokina, V. Vladimirov, {\em Journal of Physics A
}, {\bf 26} (1993),  7125--7135.
\bibitem{makar1}
A. Makarenko: {\em Control and Cybernetics}, {\bf  25}  621--630
(1996).
\bibitem{peerlart}
{\it Peerlings R.H.J.  Geers M.G.D., dr Borst R. et al} A critical
comparison of nonlocal and gradient-enhanced softening
continua//Int. J. of Solids and Structure. -- 2001. -- Vol. 38. --
P.~7723--7746.
\bibitem{danvengr}
Danilenko V.A., Belinskij I.V., Vengrovich D.B., On the
Peculiarities of the Wave Processes Connected with the Structure
of  Geophysical Media,  {\em Doklady Acad. Sci. of Ukraine\/} {\bf
12} (1996)  124--129.
\bibitem{vakhkul}
V.A. Vakhnenko  and   V.Kulich, {\em Averaged Equations of Wave
Dynamics of the Periodic Relaxing Media}, in Boundary Value Problems
(Ed. by V.A. Danylenko), Naukova Dumka, Kiev, 1990, 118--120.
\bibitem{rudyak}
V. Ya. Rudyak, {\em Statistical Theory of Dissipative Processes in
Gases and Liquids}, Nauka Publ., Novosibirsk, 1987 (in Russian).
\bibitem{zubti}
Zubarev D., Tishchenko S., {\em Physica}, {\bf 50} (1972), 285--304.
\bibitem{vladsid}
V. Vladimirov, V. Sidorets, {\em Proceedings of the Second Int.
Conference "Symmetry in Nonlinear Mathematical
 Physics" (Memorial Prof. W. Fushchych Conference July 7--13
1997, Kyiv, Ukraine), Vol. 2}, Transactions of the Institute of
Mathematics NAS of Ukraine, Kyiv, 1997, 409--417.
\bibitem{vladsidskur}
V. Vladimirov, V. Sidorets, S. Skurativsky, {\em Rep. Math.
Physics}, {\bf 44} (1999), 238-246.
\bibitem{vsandan2}
V.A. Danylenko and V.A. Vladimirov, {\em  Control and Cybernetics},
{\bf 25} (1996), 569-581.
\bibitem{dandan}
V.A. Danylenko, T.B. Danevych, O.S. Makarenko, S.I. Skurativsky and
V.A. Vladimirov, {\em Self-organization in nonlocal non-equilibrium
media}, Naukova Dumka Publ., Kyiv, 2011 (to appear).
\bibitem{baikov}
N.H. Ibragimov (ed.), {\em CRC Handbook of Lie Group Analysis of
Differential Equations, Vol. 2}, CRC Press, Boca Raton, FL, 1995.
\bibitem{wig}
S. Wiggins, {\em Global Bifurcations and Chaos: Analytical Methods},
Springer, New York, 1988.
\bibitem{meln}
V.K. Melnikov, {\em Trans Moscow Math. Soc.,} {\bf 12} (1963),
156-221.
\bibitem{G-H}
J.Guckenheimer and P.Holmes, {\em
Nonlinear Oscillations, Dynamical Systems and Bifurcations of
Vector Fields\/}, Springer, New York 1987.






\end{thebibliography}
\end{document}